\documentclass[twocolumn,twoside]{article}
\usepackage{graphicx}
\newcommand{\hb}{\\ \hspace*{2ex}}

\begin{document}
\title{SOME STATISTICAL PICTURE \hb
 OF MAGNETIC CP STARS EVOLUTION }
\author{V.F.\,Gopka$^1$, O.M.\,Ulyanov$^{2}$, S.M.\,Andrievsky$^1$,  A.V.\,Shavrina$^3$,
 V.A.\,Yushchenko$^1$\\[2mm]
\begin{tabular}{l}
 $^1$ Department of Astronomy, Odessa National University\hb
 T.G.Shevchenko Park, Odessa, 65014, Ukraine, {\em gopka.vera@mail.ru}\\
 $^2$ Institute of Radio Astronomy of NASU, \hb
  Chervonoprapona str. 4, Kharkov, 61002, Ukraine, {\em oulyanov@rian.kharkov.ua}\\
 $^3$ Main Astronomical Observatory of NASU,  \hb Zabolotnogo
str. 27, Kyiv, 03680, Ukraine\\[2mm]
\end{tabular}
}
\date{}
\maketitle

ABSTRACT. We discuss some  statistical results on the evolution of
magnetic CP stars in the framework  of the supposition about their
binary nature.


{\bf 1. Introduction}\\[1mm]
It is well known that CP stars of upper main sequence can be divided on magnetic chemically peculiar stars
(MCP stars or Bp-Ap stars) and non-magnetic (Hg-Mn stars and Am-Fm stars). Hg-Mn stars have temperatures more
than 10 000 K, Am-Fm stars are cooler than 10 000 K. Some authors concluded that there exists a relationship
between Hm-Mn stars and Am-Fm star (Adelman, Adelman A.S \& Pintado, 2003).

MCP stars are overlapped with Hg-Mn stars in the region of the hot temperatures of HR diagram and  with
Am-Fm  stars in the region of the cooler temperatures. Evolutionary state of chemically peculiar (CP) stars
of upper main sequence is the subject of some working hypotheses and numerous debates. The new observational
facts on these stars cause more and more questions. Some reviews of specialists include the list of unsolved
problems concerning this problem. Thus, it should be stated that origin of CP stars is not completely understood
at present.

The MCP stars show more complicated phenomena among CP stars
(Rudiger \& Scholz, 1988). Now we do not have a hypothesis that
could explain an origin of anomalies of chemical abundances and
the kinematics of MCP stars, as well as origin of their magnetic
field. Some characteristics of MCP stars show that they do not
support old ideas about chemical evolution. Some existing
inexplicable facts (Gopka et al., 2004, Gopka et al., 2006) force
us to suppose that MCP stars are binary systems consisting two
intermediate-mass stars experienced mass transfer between the
originally less massive star and its more massive companion in the
state of pre-supernova explosion with mass near to 8$M_{\odot}$.
As a result, the MCP star is influenced by the supernova remnant
(neutron star, NS) and some its properties are formed under this
influence (Gopka, Ulyanov \& Andrievsky, 2008a,b). Such a model is
supported by the results of the numerous investigations of MCP
stars from IR to X-rays observation. The evolutionary change of
the MCP star properties with mass reflects the change of the
system's mass ratio (both for visible and invisible neutron star
companion, Gopka, Ulyanov, Yushchenko et al., 2010).\\[2mm]

{\bf 2. On the increasing of rotational period of the low-mass MCP
stars in the framework of model MCP star binarity}\\[1mm]
 For some
MCP stars an intensive mass-loss from magnetic poles are known. As
an example, for the helium-strong stars $\sigma$~Ori~E and
HD~37017 Drake et al. (1994) estimated the mass-loss rate near
10$^{-9}$ $M_{\odot}$ per year with observed significant outflow
velocity of about 600$~km~s^{-1}$. Such a phenomenon can be easily
understood in our phenomenological model of MCP stars (see Fig
1.). Important consequence of this conclusion is supported by
statistical results of an increasing of the rotational period for
low-mass MCP stars.

\begin{figure}  
\resizebox{7cm}{!}
{\includegraphics[width=88mm]{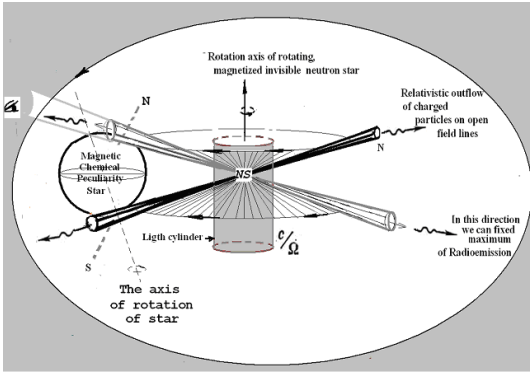}}\label{hh1}
\caption{Schematic model of MCP star as a binary system with
neutron star companion.}
\end{figure}

Fedorova (1997) investigated the qualitative changes of the
low-mass X-ray binary evolution. When the matter is accreted by
the low-mass star, the hard radiation
 occurs.
 Fedorova is obtained the theoretical dependence of the size semi-major axis from the mass of
companion for the X-ray binary (Fedorova 1997).  The some part of
the matter MCP star falls on to NS, and the some part matter is
taken away by the magnetic star wind for the case of close binary
system, into which enters MCP star and NS.
 This case is in the base of our assumption where the
donor is MCP star and acceptor is NS.

When so-called Jeans mode conditions are realized in the binary
system, the change of the system semi-major axis ($A_{m}$) due to
the mas-loss is given by the follow equation:

\begin{equation}
\label{eq1} {(\frac{dA_{m}}{dt})_{los} =  \frac{- A_{m}}{(M_{star}
+ M_{ns})} \cdot (\frac{dM_{star}}{dt})_{los}},
\end{equation}

where $A_{m}$ is the semi-major axis, $M_{star}$ is the mass of
the NS companion, $M_{ns}$ is the neutron star mass.\\

Integration of this equation produces so-called Jeans invariant:

\begin{equation}
\label{eq1} {A_{m} \cdot (M_{star} + M_{ns}) = const}.
\end{equation}

 We simulated the dependence of the semi-major
axis upon the mass of MCP star in an interval from 8$M_{\odot}$ to
 1.6$M_{\odot}$. Than were founded that semi-major axis for dual stars with the MCP star
masses in the range 1.6$M_{\odot}$ - 8$M_{\odot}$ and average NS
mass near $1.35M_{\odot}$ is grow.
 The  semi-major axis of the close binary system increases more rapidly
for the MCP stars in combination with lower mass neutron star (Fig. 2).

 This qualitatively confirms the statistical dependence obtained by Kochukhov and Bagnulo
 according to which the rotation braking (increasing of the rotational period) takes place
 in the range of small masses of the star-companion.

 The observation data from ATNF pulsar catalogue we are used also
(atnf.csiro.au).
 Among 1879 pulsars that are present in the ATNF pulsar catalogue
 only 141 PSRs enter as companions into binary systems. This are systems
 with  more number of low-mass stars. Only 81 objects having
  estimation of mass of more than 0.2$M_{\odot}$.\

For the low-masses star companion, when $M_{star} << M_{ns}$, the
semi-major axis of the system does not increase even in the case
of a strong stellar wind from the donor (Fig. 2).

 We  give the orbital period distribution as a function of the mass of the
 star-companion on Fig. 3.

\begin{figure} 
\resizebox{7cm}{!}
{\includegraphics[width=88mm]{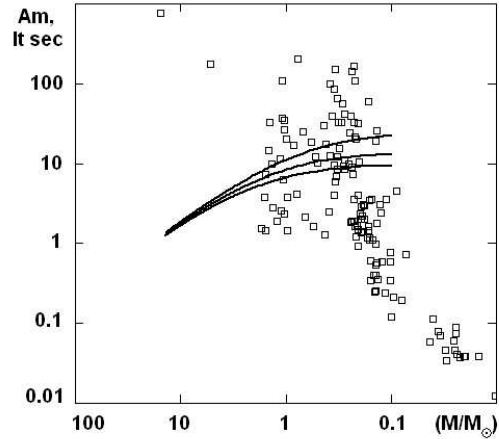}}\label{hh2}       
\caption{Distribution of the semi-major axis of the binary system
($A_{m}$) in the light second units depending on the median mass
of the neutron star companion. Solid lines illustrate qualitative
behavior under the Jeans mode conditions corresponding to
$0.8M_{\odot}, 1.4M_{\odot}, 2.0M_{\odot}$ of neutron star mass
(from the top to bottom respectively).}
\end{figure}

\begin{figure}
\resizebox{7cm}{!}
{\includegraphics[width=88mm]{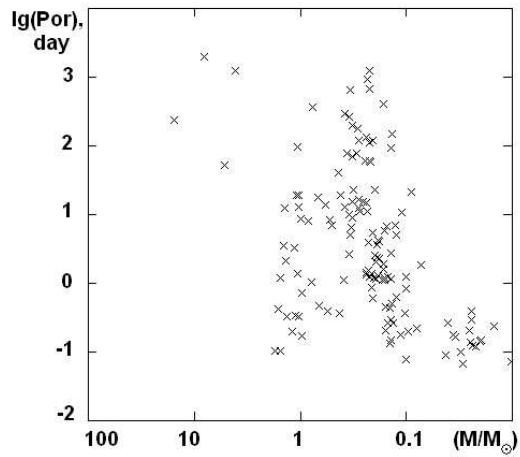}}\label{hh3}
\caption{ Distribution of the orbital periods ($P_{or}$) as a
function on companion median mass ($M/M_{\odot}$).}
\end{figure}

It is possible to assume that in the row of the binary systems in
the range of star masses from  0.3$M_{\odot}$ to 2$M_{\odot}$ the
growth of the orbital period can occur (Fig. 3.). Unfortunately,
in ATNF catalog is observed the deficit of pulsar companions in
the mass range from 2$M_{\odot}$ up to 8$M_{\odot}$ (it is MCP
mass range). It is not give the possibility to make a complete
statistical analysis.
 Figure 3 shows that the behavior with the Jeans mode condition
corresponds to another range of masses of the star-companion:
 $M_{star} \in [0.3M_{\odot}, 2.0M_{\odot}]$.\\[2mm]

{\bf 3. Conclusion}\\[1mm] We have adopted the model of MCP stars
as a close binary system with undetected neutron star as a
companion star. This model explains many properties of MCP stars,
in particular: secular decrease of the semi-major axis of the
binary systems, statistical distribution of the MCP star orbital
periods as a function on their masses, and some others.\\[2mm]

{\bf References\\[1mm]}
 Adelman S.J., Adelman A.S., Pintado O.J.: 2003, {\it A$\&$A, }{\bf 337}, 2267.\\
 Drake S.A., Lynsky J.L., Bookbinder J.A.: 1994, {\it Ap.J, }{\bf 108}, 2203.\\
 Fedorova A.V.: 1997, {\it Binary systems,} Moskow,  179.\\
 Gopka V., Yushchenko A, Shavrina A., et al. : 2004 {\it AIP Conf. Proc., }{\bf224}, 119.\\
 Gopka V., Yushchenko A, Goriely S., et al. : 2006 {\it AIP Conf. Proc., }{\bf843}, 389.\\
 Gopka V.F., Ulyanov O.M., Andrievsky S.M. : 2008, {\it AIP Conf. Proc., }{\bf 1016}, 460.\\
 Gopka V.F., Ulyanov O.M., Andrievsky S.M. : 2008, {\it KFNT, }{\bf 24}, 50.\\
 Gopka V.F., Yushchenko A.V., Yushchenko V.A., et al. : 2008, {\it KFNT, }{\bf 24}, 43.\\
 Gopka V.F., Ulyanov O.M., Yushchenko A.V., et al. : 2010, {\it AIP Conf. Proc., }{\bf 1269}, 454.\\
 Kochukhov O., Bagnulo S. : 2006, {\it A$\&$A, }{\bf 450}, 763.\\
 Rudiger G. and Scholz G. : 1988,  {\it  Astron. Nachr., }{\bf 309}, 181.\\
 http://www.atnf.csiro.au/research/pulsar/psrcat/\\

\end{document}